\documentclass[pra,nofootinbib]{revtex4}

\usepackage{graphicx}
\usepackage{amssymb}
\usepackage{amsmath}
\usepackage{amsfonts}
\usepackage{epstopdf}
\usepackage{epsfig}
\usepackage{wrapfig}
\usepackage{color}

\begin{document}

\renewcommand{\vec}[1]{\mathbf{#1}}	

\title{Markovian master equation and thermodynamics of a two-level system in a strong laser field}
\author{Krzysztof Szczygielski$^{(1)}$}
\author{David Gelbwaser-Klimovsky$^{(2)}$}
\author{Robert Alicki$^{(1)}$}
\affiliation{$^{(1)}$Institute of Theoretical Physics and Astrophysics, University of
Gda\'nsk, Gda\'nsk, Poland }
\affiliation{$^{(2)}$ Weizmann Institute of Science, Rehovot, Israel}
\begin{abstract}
The recently developed technique combining the weak-coupling limit with the Floquet formalism is applied to a model of a two-level atom driven by a strong laser field and weakly coupled to heat baths. First, the case of a single electromagnetic bath at zero temperature is discussed and the formula for resonance fluorescence is derived. The expression describes the well-known Mollow triplet, but its details differ from the standard ones based on additional simplifying assumptions. The second example describes the case of two thermal reservoirs: an electromagnetic one at finite temperature and the second dephasing one, which can be realized as a phononic or buffer gas reservoir. It is shown using the developed thermodynamical approach that the latter system can work in two regimes depending on the detuning sign: a heat pump transporting heat from the dephasing reservoir to an electromagnetic bath or heating both, always at the expense of work supplied by the laser field.

\end{abstract}

\maketitle
\section{Introduction}
The model of a two-level system (TLS)  driven by a classical electromagnetic field - laser radiation and subject to dissipation/decoherence effects is one of the most studied models in quantum optics and related fields. The basic physical phenomena - Rabi oscillations and the related appearance of Mollow triplet in the fluorescence spectrum - can be explained by the simplest version of such  models \cite{Sargent:2007}. It is based on the Markovian master equation for the reduced density matrix of the TLS which could be written as ($\hbar = k_B =1$)
\begin{equation}
{\frac{d\rho(t) }{dt}}=-i[H(t) , \rho(t)]+\mathcal{L}\rho(t)
\label{SME}
\end{equation}
where $H(t)$ is the Hamiltonian of the TLS including the driving harmonic field
\begin{equation}
H(t) = \frac{1}{2}\omega_0 \sigma^3 + g \bigl(\sigma^- e^{i\Omega t} +  \sigma^+ e^{-i\Omega t}\bigr)
\label{Ham}
\end{equation}
and $\mathcal{L}$ the phenomenological generator of Lindblad-Gorini-Kossakowski-Sudarshan (LGKS) form
\begin{equation}
\mathcal{L}\rho = \frac{\gamma_{\downarrow}}{2}\bigl([\sigma^-, \rho  \sigma^+] + [\sigma^- \rho , \sigma^+]\bigr) +\frac{\gamma_{\uparrow}}{2}\bigl([\sigma^+, \rho  \sigma^-] + [\sigma^+ \rho , \sigma^-]\bigr) - \frac{\delta}{2}[\sigma_3,[\sigma_3,\rho]]
\label{SME1}
\end{equation}
with phenomenological damping $(\gamma_{\downarrow})$, pumping $(\gamma_{\uparrow})$, and dephasing $(\delta)$ rates.
\par
It long has been noticed that the above equation is not consistent with the second law of thermodynamics
and also does not reproduce experimental results well for the strong driving case (see Ref. \cite{Geva:1995} for the discussion and numerous references). The reason is that the dissipation and decoherence mechanisms  are not independent of the Hamiltonian dynamics, which is now time dependent and, hence, one should expect that the simple phenomenological generator (\ref{SME1}) should be replaced by a time-dependent superoperator. There exist several examples in the literature addressing this problem for different physical systems like atoms, NMR, quantum dots, and superconducting qubits. The authors derive  master equations
using various approximation schemes \cite{{Kamleitner:2011}}. In the present paper we use a new systematic approach which combines the Davies idea of the weak-coupling limit \cite{Davies:1974} with the Floquet theory of periodically driven quantum systems. This approach has been outlined in Ref. \cite{Alicki:2006} and used for different models of quantum engines and refrigerators \cite{Levy:2012}. As already noted, the mathematical structure of the  obtained (time-inhomogeneous)  Markovian master equation is consistent with the second-law of thermodynamics and allows us to define unambiguously heat currents and stationary power when applied to models of quantum machines. \\
The motivation to improve the existing formalisms, as well as for the TLS dynamics governed by the Hamiltonian (\ref{Ham}) (which has not been studied within this new framework so far), is as follows:\\
(1) The experiments concerning control of microscopic quantum objects like single atoms, quantum dots, or nanoscopic superconducting devices become so precise that the commonly used approximations may be insufficient.\\
(2) Recently, thermodynamics of microscopic quantum machines has been studied intensively and it is important that the theoretical formalism does not introduce spurious inconsistencies with thermodynamics which are entirely due to the chosen approximation schemes (see Ref. \cite{Alicki:1979} for a sample of references).\\
(3) The advocated formalism produces rather simple canonical quantum master equations, which reduce to simple kinetic equations for populations in a proper orthonormal basis. This allows study analytically various new designs of microscopic quantum machines.
\par
We do not repeat here the derivation of the master equation for periodically driven quantum open systems, which can be found in Ref. \cite{Alicki:2006},  but present only its main ingredients for its construction and basic properties without proofs (Secs. II and III). We then discuss the case of TLS with the Hamiltonian (\ref{Ham}) applied to two particular models. The first one corresponds to the \emph{resonance fluorescence} phenomenon, including detuning effects. Although the result qualitatively reproduces the well-known Mollow triplet, there are quantitative differences  in comparison with the existing approximation schemes which, in principle, should be detectable.  The second model presents a microscopic heat pump based on the driven TLS (atom, quantum dot, nanoparticle, etc.)  which illustrates a new coherent mechanism of energy transfer among TLS, an electromagnetic field, and a generic dephasing bath, which involves Rabi oscillations.

\section{Master equation for periodically driven open system}
A ``small" quantum system (i.e., described by a Hamiltonian with a discrete spectrum) is driven by an external periodic field
and weakly coupled to an environment. The environment is assumed to be a ``large" quantum system which can be treated as a system with continuous spectrum satisfying all standard ergodic conditions. They allow us to apply the weak-coupling limit procedure to derive the Markovian master equation for the reduced density matrix of the small system. The reversible effect of driving combined with the Hamiltonian corrections caused by the interaction with environment (``Lamb shift") is described by the physical (renormalized) periodic Hamiltonian $H(t)= H(t + \tau)$. We assume the following form of the system-environment coupling:
\begin{equation}
H_{int} = S\otimes F\ ,\ \langle F\rangle_E =0.
\label{coupling}
\end{equation}
where the environment is assumed to be at the stationary state and $\langle \cdots\rangle_E $ denotes the associated quantum average. In particular, in the weak-coupling Markovian approximation the influence of the environment is completely characterized by its spectral density defined as
\begin{equation}
G(\omega)= \int_{-\infty}^{+\infty} e^{i\omega t}\langle F(t)F\rangle_E dt \geq 0,
\label{spectral}
\end{equation}
where $F(t)$ is an environmental observable evolving under the Heisenberg evolution governed by the environmental Hamiltonian. In the next step of construction we decompose the time-dependent (in the Heisenberg picture) coupling operator $S$ into the Fourier components
\begin{equation}
S(t) = U^{\dagger}(t) S U(t) = \sum_{q\in \mathbf{Z}}\sum_{\{\bar{\omega}\}} S_q(\bar{\omega})e^{-i (\bar{\omega}+ q\Omega)t}, \quad \Omega = \frac{2\pi}{\tau}
\label{fourier}
\end{equation}
where $U(t)$ is a unitary propagator defined by the time-ordered exponential
\begin{equation}
U(t) = \mathbf{T}\exp\Bigl\{-i\int_0^{t} H(s) ds\Bigr\}.
\label{propagator}
\end{equation}
The detailed structure of (\ref{fourier}) is a consequence of the Floquet theory. Namely, one can uniquely define the \emph{averaged Hamiltonian} and its spectral decomposition,
\begin{equation}
U(\tau) = e^{-i\bar{H}\tau}\ ,\ \bar{H} = \sum_k \bar{\epsilon}_k |k\rangle\langle k|
\label{propagator2}
\end{equation}
The differences of quasienergies $\{\bar{\epsilon}_k - \bar{\epsilon}_l\}$ form a set of \emph{Bohr quasifrequencies}
denoted by $\{\bar{\omega}\}$ and the operators  $S_q(\bar{\omega})$ satisfy the relation
\begin{equation}
[\bar{H} ,S_q(\bar{\omega})] = -\bar{\omega} S_q(\bar{\omega})\ ,\  S_{-q}(-\bar{\omega}) = {S^{\dagger}}_q(\bar{\omega}).
\label{propagator1}
\end{equation}
Those \emph{transition operators} called also \emph{Lindblad operators} describe the transitions between the levels of the averaged Hamiltonian due to the interaction with environment and assisted by the energy exchange $q\Omega$ with an external source of periodic perturbations. They are the main ingredients of the following LGKS generator describing the irreversible evolution of an open system in the interaction picture. This generator is  a sum of \emph{local generators} each of the LGKS type
\begin{equation}
\mathcal{L} = \sum_{q\in \mathbf{Z}}\sum_{\{\bar{\omega}\geq 0\}} \mathcal{L}_{q\bar{\omega}}
\label{generator}
\end{equation}
\begin{equation}
\mathcal{L}_{q\bar{\omega}}\rho = \frac{1}{2}\Bigl\{G(\bar{\omega}+ q\Omega)\bigl([S_q(\bar{\omega})\rho, S^{\dagger}_q(\bar{\omega})] + [S_q(\bar{\omega}), \rho S^{\dagger}_q(\bar{\omega})]\bigr) + G(-\bar{\omega}- q\Omega)\bigl([S^{\dagger}_q(\bar{\omega})\rho, S_q(\bar{\omega})] + [S^{\dagger}_q(\bar{\omega}), \rho S_q(\bar{\omega})]\bigr)\Bigr\}
\label{generator_loc}
\end{equation}
Notice that this construction can be easily generalized to  more complicated interaction Hamiltonians  of the form
$H_{int}= \sum_{j} S^{j}\otimes F_{j}$. Moreover, in the case when an environment consists of independent subsystems labeled by $j$, the construction is additive, leading to the generator
\begin{equation}
\mathcal{L} = \sum_{j}\sum_{q\in \mathbf{Z}}\sum_{\{\bar{\omega}\geq 0\}} \mathcal{L}^{j}_{q\bar{\omega}}\ \mathrm{with}\  G(\omega)\mapsto G^{j}(\omega)\ ,\  S_q(\bar{\omega}) \mapsto S^{j}_q(\bar{\omega}).
\label{generator1}
\end{equation}
Finally, we can write down the Markovian master equation in the Schr\"odinger picture
\begin{equation}
{\frac{d\rho(t) }{dt}}=-i[H(t) , \rho(t)]+\mathcal{L}(t)\rho(t)
\label{ME_gen}
\end{equation}
where
\begin{equation}
\mathcal{L}(t)\rho = \mathcal{U}(t)\mathcal{L}\bigl( \mathcal{U}^{-1}(t)\rho )\ ,\ \mathcal{U}(t)\rho \equiv U(t) \rho U^{\dagger}(t)
\label{ME_gen1}
\end{equation}
The properties (\ref{propagator}),  (\ref{propagator1}), (\ref{generator}), and  (\ref{generator1}) imply that the solution of the master equation (\ref{ME_gen}) can be written in a factorized form,
\begin{equation}
\rho(t) = \mathcal{U}(t)e^{t\mathcal{L}}\rho(0).
\label{factor}
\end{equation}
The decomposition (\ref{factor}) is very useful, because it allows us to find the long-time evolution of the system. Namely, if $\mathcal{L}$ possesses a unique stationary state $\tilde{\rho}$ (i.e.,
$\mathcal{L}\tilde{\rho}=0$), then  for $t\to\infty$,  $e^{t\mathcal{L}}\rho(0) \to \tilde{\rho}$ and the system asymptotically reaches the periodic limit cycle (which for some models is degenerated to a single stationary state)
described by the formula
\begin{equation}
\tilde{\rho}(t) = \mathcal{U}(t)\tilde{\rho}= \tilde{\rho}(t+ \tau).
\label{limit}
\end{equation}
Periodicity follows from the commutation of $\tilde{\rho}$ with $U(\tau)$. In the steady-state regime, only the properties of interaction picture generators $\mathcal{L}^j_{q\bar{\omega}} $ matter. In particular, those generators preserve the diagonal (in $\bar{H}$ basis) elements of the density matrix, which is a consequence of  \eqref{propagator1}. Therefore, the calculations of basic quantities for the steady state involve only the corresponding ``classical" Pauli master equations. This property simplifies the analysis of various models and often allows for analytical solutions.
\section{Thermodynamical properties}
The rich structure and nontrivial mathematical properties of the master equations discussed above allow us to define consistently heat currents and stationary power, and derive rigorously the second law of thermodynamics for driven open quantum systems. \\
\subsection{Heat currents and power }
The crucial property of the master equation \eqref{ME_gen} is its additivity with respect to  the independent system-reservoir couplings labeled by $j$, the Bohr quasifrequencies $\{\bar{\omega}\geq 0\}$, and the harmonics $\{q\Omega\}$ of the periodic perturbation.
The similar additivity is expected for the heat current flowing through the system, which should be a sum of \emph{local time-dependent heat currents} $\mathcal{J}^j_{q\bar{\omega}}(t)$. They are defined as
\begin{equation}
\mathcal{J}^j_{q\bar{\omega}}(t) = \frac{\bar{\omega} + q\Omega}{\bar{\omega}}\mathrm{Tr}\bigl[( \mathcal{L}^j_{q\bar{\omega}}(t)\rho(t))\bar{H}(t)\bigr]
\label{curr_loc}
\end{equation}
where
\begin{equation}
\bar{H}(t)= U^{\dagger}(t) \bar{H} U(t).
\label{curr_loc1}
\end{equation}
The formula (\ref{curr_loc}), albeit a bit complicated, possesses a clear physical interpretation.  The temporal rate of the energy change entirely due to the $j$-th component of the bath, $q$-th harmonic of the external periodic perturbation, and the transitions between the levels of the temporal averaged Hamiltonian $\bar{H}(t)$ separated by the Bohr quasifrequency $\bar{\omega}$ is given by $\mathrm{Tr}\bigl[(\mathcal{L}^j_{q\bar{\omega}}(t)\rho(t))\bar{H}(t)\bigr]$.
The ``renormalizing" coefficient $(\bar{\omega}+ q\Omega)/\bar{\omega}$ takes into account that the system energy change $\bar{\omega}$ is accompanied by the energy exchange $q\Omega$ with an external source of the periodic perturbation.
\par
We assume now that the $j$-th reservoir is a thermal bath at the equilibrium state characterized by the temperature $T^j$ and the Kubo-Martin-Schwinger
condition for an arbitrary choice of the bath observable $F$ in \eqref{spectral}
\begin{equation}
G^j(-\omega) = e^{-\omega/T^j} G(\omega) .
\label{temp}
\end{equation}
The relation \eqref{temp} implies the existence of Gibbs-like time-dependent stationary state $\tilde{\rho}^j_{q\bar{\omega}}(t)$ for any local time-dependent Schr\"{o}dinger picture generator $\mathcal{L}^j_{q\bar{\omega}}(t)$ and similarly for a constant interaction picture one,
\begin{equation}
\mathcal{L}^j_{q\bar{\omega}}(t)\tilde{\rho}^j_{q\bar{\omega}}(t)= 0,\quad \mathcal{L}^j_{q\bar{\omega}}\tilde{\rho}^j_{q\bar{\omega}} = 0
\label{stat_loc}
\end{equation}
\begin{equation}
\tilde{\rho}^j_{q\bar{\omega}}(t)= U(t)\tilde{\rho}^j_{q\bar{\omega}}U^{\dagger}(t),\quad \tilde{\rho}^j_{q\bar{\omega}} = \frac{\exp\Bigl\{-\frac{(\bar{\omega}+q\Omega)\bar{H}}{\bar{\omega}T^j}\Bigr\}}{\mathrm{Tr}\exp\Bigl\{-\frac{(\bar{\omega}+q\Omega)\bar{H}}{\bar{\omega}T^j}\Bigr\}} .
\label{stat_loc1}
\end{equation}
The the local current then can be rewritten in the following form:
\begin{equation}
\mathcal{J}^j_{q\bar{\omega}}(t)= - T^j\mathrm{Tr}\bigl[( \mathcal{L}^j_{q\bar{\omega}}(t)\rho(t))\ln \tilde{\rho}^j_{q\bar{\omega}}(t)\bigr].
\label{curr_loc2}
\end{equation}
and the heat current flowing from the $j$-th  bath is defined as
\begin{equation}
\mathcal{J}^j(t) =\sum_{q\in \mathbf{Z}}\sum_{\{\bar{\omega}\geq 0\}} \mathcal{J}^j_{q\bar{\omega}}(t) .
\label{j-current}
\end{equation}
\subsection{The second law}
The most important result which follows from the theory presented above is the \emph{second law of thermodynamics} written as a positive entropy production,
\begin{equation}
\frac{d}{dt} S\bigl(\rho(t)\bigr) -  \sum_j \frac{1}{T^j}\mathcal{J}^j(t)\geq 0
\label{IIlaw}
\end{equation}
where $S(\rho)=-\mathrm{Tr}(\rho\ln\rho)$ and the inequality \eqref{IIlaw} is a consequence of \eqref{curr_loc2} and the Spohn inequality
$\mathrm{Tr}\bigl[(\mathcal{L}\rho)(\ln\rho  - \ln\tilde{\rho}\bigr])\leq 0$ is valid for any LGKS generator $\mathcal{L}$
and its stationary state $\tilde{\rho}$ \cite{Spohn:1978}.\\
\par
We are often interested in the steady-state (limit cycle) properties. In this case, the formulas simplify, whereby the local heat currents and the entropy become time independent. Using the following notation for a stationary local current:
\begin{equation}
\tilde{\mathcal{J}}^j = \sum_{q\in \mathbf{Z}}\sum_{\{\bar{\omega}\geq 0\}}\frac{\bar{\omega} + q\Omega}{\bar{\omega}}\mathrm{Tr}\bigl[( \mathcal{L}^j_{q\bar{\omega}}\tilde{\rho})\bar{H}\bigr]
\label{curr_stat}
\end{equation}
one can write the second law for the stationary regime as
\begin{equation}
\sum_j \frac{\tilde{\mathcal{J}}^j}{T^j}\leq 0 .
\label{IIlaw_stat}
\end{equation}
Due to the energy conservation, one can define the stationary power of the external periodic force executed on the system as
\begin{equation}
\tilde{\mathcal{P}}= - \sum_j \tilde{\mathcal{J}}^j.
\label{power}
\end{equation}
One should note that due to the properties \eqref{propagator1}, \eqref{generator_loc}, and \eqref{stat_loc1}, the interaction picture local generators $\mathcal{L}^j_{q\bar{\omega}}$
transform the diagonal part [in $\bar{H}$ basis \eqref{propagator2}] of the density matrix into a diagonal one and all stationary states $\tilde{\rho}^j_{q\bar{\omega}},\tilde{\rho}$ are also diagonal, which drastically simplifies all computations for particular examples.

\section{Examples}
We consider  models of a TLS system with the Hamiltonian (\ref{Ham}) coupled to (i) a single zero-temperature bath (electromagnetic field) and (ii) two heat baths at different temperatures. The first model provides a description of  fluorescence in a vacuum and will be used to compare our approach with the standard theory based on the Eqs. (\ref{SME})-(\ref{SME1}). The second one is expected to describe a microscopic heat pump powered by laser radiation.

\subsection{Preliminaries: \textit{Transition operators} for couplings between TLS and environment}

The propagator (\ref{propagator}) for the Hamiltonian (\ref{Ham}) can be written as a product of two one-parameter unitary groups,
\begin{equation}
U(t)=U_1 (t) U_2 (t),\qquad U_{1}(t) = \mbox{exp}\left(-\frac{1}{2}it\Omega\sigma^3\right) , \, \, U_{2}(t) = \mbox{exp}\left\{-it\left(\frac{1}{2}\Delta \sigma^{3} + g\sigma^1\right)\right\} .
\label{TLS_prop}	
\end{equation}
In order to prove (\ref{TLS_prop}) it is enough to differentiate two matrix-valued functions--$U(t)$ given by \eqref{propagator} and $U_{1}(t)U_{2}(t)$--to show that they satisfy the same differential equation. The mean Hamiltonian $\bar{H}$ reads
\begin{equation}\label{meanHamiltonian}
  \bar{H} = \frac{1}{2}\Delta \sigma^{3} + g\sigma^1,
\end{equation}
where $\Delta = \omega_0 - \Omega$ is the detuning parameter.
Next we diagonalize this Hamiltonian,  such that $\bar{H} = \epsilon_{1} \left| \phi_{1} \right\rangle \left\langle \phi_{1} \right| + \epsilon_{2} \left| \phi_{2} \right\rangle \left\langle \phi_{2} \right|$, with
\begin{subequations}\label{eigenProblemHbar}
  \begin{equation}
    \epsilon_{1} = -\frac{1}{2} (\Omega - \Omega_{R}), \qquad \epsilon_{2} = -\frac{1}{2} (\Omega + \Omega_{R}),
  \end{equation}
  \begin{equation}
    \left|\phi_{1}\right\rangle = \frac{1}{\sqrt{4g^2 + (\Delta + \Omega_{R})^2}} \left( \begin{array}{c} \Delta + \Omega_{R} \\ 2g\end{array} \right), \qquad
    \left|\phi_{2}\right\rangle = \frac{1}{\sqrt{4g^2 + (\Delta - \Omega_{R})^2}} \left( \begin{array}{c} \Delta - \Omega_{R} \\ 2g\end{array} \right),
  \end{equation}
\end{subequations}
where $\Omega_R = \sqrt{4g^2 + \Delta^2}$ is the Rabi frequency (we assume always  $\Omega_R \leq \Omega$). Obviously, the vectors $\{\phi_{1},\,\phi_{2}\}$ form an orthonormal basis and, from now on, all matrix representations of appearing operators shall be explicitly written in terms of this basis. We choose the following convention for those basis vectors:\\
\begin{equation}
    \left|\phi_{1}\right\rangle = \left( \begin{array}{c} 1 \\ 0\end{array} \right), \qquad
    \left|\phi_{2}\right\rangle =  \left( \begin{array}{c} 0 \\ 1\end{array} \right).
\label{basis}
\end{equation}
First, we give the detailed structure of  transition operators for two types of TLS-environment couplings.
\\
\subsubsection{Coupling by $\sigma^1 \otimes F$ interaction Hamiltonian}

We compute now explicitly the Fourier decomposition (\ref{fourier}) of $\sigma^1(t)$. Using (\ref{TLS_prop}) we obtain
\begin{equation}
\sigma^{1} (t) = \left(\frac{\Delta}{\Omega_{R}} \cos{\Omega t} \cos{\Omega_{R} t} - \sin{\Omega t} \sin{\Omega_{R} t}\right)\sigma^{1} - \left( \cos{\Omega_{R} t} \sin{\Omega t} + \frac{\Delta}{\Omega_{R}} \cos{\Omega t} \sin{\Omega_{R} t} \right)\sigma^{2} + \left( \frac{2g}{\Omega_{R}} \cos{\Omega t} \right)\sigma^{3},
\label{fou_sig_1}
\end{equation}
and then its matrix representation reads [in the new basis \eqref{basis}]
\begin{equation}\label{Sigma1Int}
	\sigma^{1}(t) = \frac{1}{\Omega_{R}} \left(
\begin{array}{cc}
 2g\cos{\Omega t} & e^{i \Omega_R t} \left( \Delta \cos{\Omega t} + i \Omega_{R} \sin{\Omega t} \right) \\
 e^{-i \Omega_R t} \left( \Delta \cos{\Omega t} - i \Omega_{R} \sin{\Omega t} \right) & -2g\cos{\Omega t}
\end{array}
\right).
\end{equation}
By inserting exponential functions into trigonometric ones, one can find all components of the Fourier decomposition of this operator. The only positive frequencies correspond to $\bar{\omega}= 0, \pm\Omega_R$ and $q=1$ and yield the transition operators of the form
	\begin{equation}
		S^1(\Omega-\Omega_R) = \frac{1}{2 \Omega_{R}}\left(
\begin{array}{cc}
 0 & \Delta - \Omega_{R} \\
 0 & 0
\end{array}
\right), \qquad
		S^1(\Omega) = \frac{g}{\Omega_{R}} \left(
\begin{array}{cc}
 1 & 0 \\
 0 & -1
\end{array}
\right), \qquad
		S^1(\Omega+\Omega_R) = \frac{1}{2 \Omega_{R}}\left(
\begin{array}{cc}
 0 & 0 \\
 \Delta + \Omega_{R} & 0
\end{array}
\right),
	\end{equation}
	\label{Sigma1Freq}
with the symmetry $S^1(-\omega) = {S^1}(\omega)^{\dagger}$ [compare eq.(\ref{propagator1})].

\subsubsection{Coupling by $\sigma^3 \otimes F$ interaction Hamiltonian}

Just as previously, we have to perform all the necessary mathematical steps once again, this time for the matrix $\sigma^3$ to obtain the corresponding transition operators. Again,
\begin{equation}
\sigma^{3} (t)= \left(-\frac{2g}{\Omega_{R}} \cos{\Omega_R t}\right)\sigma^{1}+\left(\frac{2g}{\Omega_{R}} \sin{\Omega_R t}\right)\sigma^{2} + \frac{\Delta}{\Omega_{R}} \sigma^{3},
\label{fou_sig_2}
\end{equation}
with the matrix representation
\begin{equation}
		\sigma^{3}(t) = \frac{1}{\Omega_{R}} \left(
\begin{array}{cc}
 \Delta & -2g e^{i \Omega_R t} \\
 -2g e^{-i \Omega_R t} & -\Delta
\end{array}
\right) .
\label{Rabi}
	\end{equation}
Notice that now only two non-negative frequencies appear, 0 and the Rabi frequency $\Omega_R$, and with corresponding transition operators,
\begin{equation}
S^3(0) = \frac{\Delta}{\Omega_{R}}\left(
\begin{array}{cc}
 1 & 0 \\
 0 & -1
\end{array}
\right),
\qquad
S^3(\Omega_R) = \frac{2g}{\Omega_{R}}\left(
\begin{array}{cc}
 0 & 0 \\
 -1 & 0
\end{array}
\right).
\end{equation}
and $S^3(-\Omega_R)= S^3(\Omega_R)^{\dagger}$.
\subsection{Example 1: Resonance Fluorescence in vacuum with detuning}

As a first example, we will consider the case of fluorescence in vacuum \cite{Sargent:2007}. A TLS with periodic driving of frequency $\Omega$ described by the time-dependent Hamiltonian \eqref{Ham} is weakly coupled to an electromagnetic bath at zero temperature (vacuum). The interaction Hamiltonian between the TLS and the electromagnetic field is given under a \textit{dipole approximation} as
\begin{equation}\label{ResonanceFluorescenceInteraction}
	H_{int} =  \sigma^{1} \otimes ( a^{\dagger} (f) + a(f) ).
\end{equation}
Notice that the usual \textit{rotating wave approximation} (RWA) was \textbf{not} performed. The short-hand notation
\begin{equation}\label{CreatorAnnihilator}
a(f) = \sum_{\mu=\pm 1}\int\limits_{\mathbb{R}^3} \overline{f_{\mu}(\vec{k})} a_{\mu}(\vec{k}) \, \mbox{d}^{3} k,
\end{equation}
where $\vec{k}$ is a wave vector,  $\mu$ denotes polarization, and  $\overline{f_{\mu}(\vec{k})}$ is a properly chosen form factor used with the electromagnetic field operators satisfying
$[a_{\mu}(\vec{k}), {a_{\mu'}}^{\dagger} (\vec{k}')] = \delta_{\mu\mu'}\delta (\vec{k} - \vec{k}')$.
The Hamiltonian of electromagnetic bath is given by
\begin{equation}\label{EMhamiltonian}
	H_{R} = \sum_{\mu=\pm 1} \int\limits_{\mathbb{R}^{3}} \omega(\vec{k}) \, {a_{\mu}}^{\dagger} (\vec{k}) a_{\mu}(\vec{k}) \, \mbox{d}^{3} k, \qquad \omega(\vec{k}) = c |\vec{k}|.
\end{equation}
The standard choice of the form factor  $f_{\mu}(\vec{k})= i (2\pi)^{-\frac{3}{2}} \sqrt{c |\vec{k}| /\epsilon_0} (\vec{d} \cdot \vec{e}_{\vec{k}\mu})$, where $\vec{d}$ is the transition dipole and $\vec{e}_{\vec{k}\mu}$ determines polarization, yields the spectral density at vacuum state
\begin{equation}
	G(\omega) = \mathcal{A} \omega^3 , \qquad  \mathcal{A} = \frac{|\vec{d}|^{2}}{3\pi\epsilon_0 c^3}.
\end{equation}
\\
For convenience, let us introduce some useful notation,
\begin{equation}
\delta_0 =	\Bigl(\frac{2g}{\Omega_R}\Bigr)^2 G(\Omega), \quad \delta_{\pm} = \Bigl(\frac{ \Omega_{R} \pm \Delta}{2\Omega_R}\Bigr)^2 G(\Omega \pm \Omega_R),
\label{para_1}
\end{equation}
\begin{equation}
\gamma_1 = \delta_{-} + \delta_{+} , \quad \gamma_2 = \frac{1}{2} (\delta_{-} + \delta_{+} +\delta_0) .
\label{para_2}
\end{equation}
Performing all the necessary calculations and summing up all parts like in the prescription in \eqref{generator}, \eqref{generator_loc}, and \eqref{generator1}, one obtains the resulting semigroup generator $\mathcal{L}$ acting on the reduced density matrix $\rho^I(t)$ of TLS  in the interaction picture and, finally, the associated master equation in matrix elements [$\rho^I_{11}(t)+\rho^I_{22}(t)=1$]
\begin{equation}
		\frac{d\rho^I_{11}(t)}{dt}= -\delta_{+}\rho^I_{11}(t) + \delta_{-}\rho^I_{22}(t), \quad \frac{d\rho^I_{12}(t)}{dt}= -\gamma_2 \rho^I_{12}(t)
	 	 \label{master_flu}
	 \end{equation}
with the solution
	\begin{equation}
		\rho^I_{11}(t) = e^{-\gamma_1 t} \rho_{11} (0) + \frac{\delta_{-}}{\gamma_1} \bigl(1- e^{-\gamma_1 t}   \bigr), \quad  \rho^I_{12}(t) = e^{-\gamma_2 t}\rho_{12}(0)
	\end{equation}

\begin{figure}[!ht]
	\centering
		\includegraphics[width=1.00\textwidth]{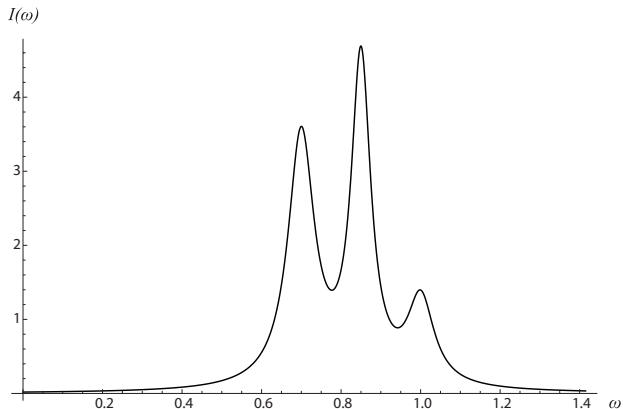}
	\caption{Power spectral density function \eqref{PowerDensityNoRWA} in a process of nonresonant fluorescence in vacuum ($\Omega = 0.85$, $\Delta = 0.01$, $g = 0.075$). The chosen ratio $g/\Omega$ and the value of $\mathcal{A}$ are very high in comparison to known physical examples in order to make visible the asymmetry of the line shape.}
	\label{fig:PowerDensityNoRWA}
\end{figure}

Using those results, we can calculate the \emph{fluorescence power spectrum}. We apply the standard formula describing the \textit{power spectral density function} of incident light scattered by a TLS coupled to an electromagnetic reservoir, which can be found, for example, in Ref. \cite{Sargent:2007}
\begin{equation}
	I(\omega) \propto \mbox{Re} \int\limits_{0}^{\infty} e^{-i\omega t} \, \mbox{tr}\left\{ \sigma^{+} \mathcal{U}(t) e^{t\mathcal{L}} (\sigma^{-} \tilde{\rho}) \right\} \, \mbox{d}t,
\end{equation}
where $\sigma^{\pm} = \frac{1}{2} (\sigma^{1} \pm i\sigma^{2})$, $\mathcal{U}(t) e^{t\mathcal{L}}$ is a quantum dynamical map in the Schr\"{o}dinger picture and $\tilde{\rho}$ stands for stationary state of semigroup generator, satisfying $\mathcal{L}\tilde{\rho} = 0$. After straightforward  calculations, one can find the formula
\begin{align}\label{PowerDensityNoRWA}
		I(\omega) \propto \left(\frac{\delta_{-}-\delta_{+}}{\gamma_1 }\right)^2 &\times \delta(\omega -\Omega )\\
	+	4\Bigl(\frac{4 g^2 \delta_{+}}{(\Omega_R - \Delta)^2\,\gamma_1 }\Bigr)\Bigl(\frac{4 g^2 \delta_{-}}{(\Omega_R + \Delta)^2\,\gamma_1 }\Bigr) &\times \frac{1}{\pi}\frac{\gamma_1}{{\gamma_1}^2+(\omega -\Omega )^2}\nonumber \\
		 +\frac{4 g^2 \delta_{+}}{(\Omega_R + \Delta)^2\,\gamma_1 } & \times \frac{1}{\pi}\frac{\gamma_2}{{\gamma_2}^2+[\omega -(\Omega -\Omega_R)]^2 }
		\nonumber \\
 +\frac{4 g^2 \delta_{-}}{(\Omega_R - \Delta)^2\,\gamma_1 } & \times \frac{1}{\pi}\frac{\gamma_2}{{\gamma_2}^2+[\omega -(\Omega +\Omega_R)]^2 },
		\nonumber
\end{align}
which clearly possesses the structure of the usual Mollow spectrum (\textit{Mollow triplet}), with Lorentzian peaks centered around frequencies $\Omega$ (central peak) and $\Omega \pm \Omega_R$ (side peaks). The Dirac delta contribution originates from possible elastic scattering of incident light without changing the energy of scattered photons. All terms are products of Lorentzians (or Dirac delta) with normalized integrals and weights describing the relative contributions to the total power of the scattered radiation. Although, qualitatively, the obtained line-shape is standard, the details of the coefficients differ from those found in the literature, in particular computed using \eqref{SME1} (see, e.g., Ref. \cite{Sargent:2007}). Our formulas contain more details of the model including, for example, the frequency dependence of the decay rates.  The plot of the obtained power spectral density function \eqref{PowerDensityNoRWA} is presented in Fig. \ref{fig:PowerDensityNoRWA}. Notice that the chosen ratio $g/\Omega$ and the value of $\mathcal{A}$ are very high in comparison to known physical examples in order to make visible the asymmetry of the line shape.

\subsection{Example 2: A microscopic heat pump driven by a laser}

As a second example, we will consider a model of a TLS, periodically driven by a laser field and weakly coupled to two heat baths: an \emph{electromagnetic (photon) bath} at the temperature $T_{e} > 0$ and a \emph{dephasing bath} at the temperature $T_d >0$. The TLS-bath interaction Hamiltonians differ for both cases: The photon bath is coupled by the Hamiltonian of the form  \eqref{ResonanceFluorescenceInteraction} (i.e., $\sigma^1 \otimes F$ -coupling) while the coupling to the dephasing bath is
of the $\sigma^3 \otimes F$ -type. The second form of coupling means that without external driving the second bath cannot cause transitions between the TLS energy levels and produces only dephasing (pure decoherence) of the TLS state.
This is typically due to the energy mismatch between high energy level splitting of the TLS and low energy excitations of the bath. The external periodic driving together with the presence of the electromagnetic bath generate energy transfer between the laser field and both heat baths mediated by the TLS. It follows from the formula \eqref{Rabi} which shows that the driven TLS is coupled to the dephasing bath through oscillations with Rabi frequency which is much lower and, hence, can match bath excitations.
\par
The following examples of physical systems satisfy the assumptions of this model:\\

(1) \emph{An optically active atom immersed in a dense buffer gas.}  The collisions with the buffer gas are elastic; hence, the condition of a pure dephasing bath is satisfied. Recently, both cooling (for red-detuned laser radiation) and heating (for blue-detuned laser radiation) have been observed using rubidium atoms in
argon buffer gas \cite{Vogl:2009}.\\

(2) \emph{A quantum dot interacting with acoustic phonons.} As the optical frequency of the dot is much higher than the Debye frequency, a  ``bottleneck effect"  appears \cite{Xin}, which suppresses the energy exchange between the dot and the phonon bath. Again, a proper detuning of the laser beam should generate cooling or heating of the phonon bath.\\

(3) \emph{A metallic nanoparticle supporting a TLS and coupled to a macromolecule.} The  molecule vibrational degrees of freedom form a dephasing baths. Recently, such hybrid systems driven by laser radiation were investigated in the context of molecular electronics \cite{Neubauer:2012}. It seems that such systems could also be used as nanomachines (heat pumps or engines).\\
\par
The quantitative analysis of the discussed model is based on the master equation and the thermodynamical formalism developed in Secs. II and III. Similarly to the previous example, we can use the formulas for transition maps and construct the interaction picture generator. We use also a short-hand notation similar to that from the previous section,
\begin{equation}
\delta_{0} =	\Bigl(\frac{2g}{\Omega_R}\Bigr)^2 G^d(\Omega_R), \quad \delta_{\pm} = \Bigl(\frac{ \Omega_{R} \pm \Delta}{2\Omega_R}\Bigr)^2 G^e(\Omega_{\pm}),\quad \Omega_{\pm}=\Omega \pm \Omega_R
\label{para_3}
\end{equation}
where $G^e(\omega )$ and $G^d(\omega )$ are spectral densities for the electromagnetic bath and dephasing bath, respectively.
\par
To obtain the expressions for steady-state heat currents we need only the following equation for the diagonal matrix
 elements in the interaction picture ($\rho^I_{22} = 1- \rho^I_{11}$):
  \begin{equation}
		\frac{d\rho^I_{11}(t)}{dt}=  -\left( \delta_{0} + e^{-\frac{\Omega_{-}}{T_{e}}} \delta_{-} + \delta_{+} \right)\rho^I_{11}(t) + \left( e^{-\frac{\Omega_{R}}{T_{d}}}\delta_{0} + \delta_{-} + e^{-\frac{\Omega_{+}}{T_{e}}} \delta_{+} \right) \rho^I_{22}(t) ,
	 	 \label{dot_eq}
	 \end{equation}
Finally, we compute heat currents for both baths using the formulas \eqref{curr_stat} and the stationary state obtained from \eqref{dot_eq}. The matrix elements of $\tilde{\rho}$ satisfy
\begin{equation}
\frac{\tilde{\rho}_{11}}{\tilde{\rho}_{22}}= \frac{e^{-\frac{\Omega_{R}}{T_{d}}}\delta_{0} + \delta_{-} + e^{-\frac{\Omega_{+}}{T_{e}}} \delta_{+}}{\delta_{0} + e^{-\frac{\Omega_{-}}{T_{e}}} \delta_{-} + \delta_{+}}
\label{stat_dot}	
\end{equation}
The formulas for the heat currents flowing from the dephasing and the electromagnetic baths read
\begin{equation}
\tilde{\mathcal{J}}^d = - \frac{\delta_{0}}{\mathcal{K}} \left[ \delta_{-} \left( 1-e^{- \left( \frac{\Omega_{R}}{T_{d}} + \frac{\Omega_{-}}{T_{e}} \right)}\right) + \delta_{+} \left( e^{-\frac{\Omega_{+}}{T_{em}}} - e^{-\frac{\Omega_{R}}{T_{d}}} \right) \right]
\label{current_dot1}	
\end{equation}
\begin{equation}
\tilde{\mathcal{J}}^{e} = \frac{1}{\Omega_R\mathcal{K}} \left[ 2\left( e^{-\frac{2\Omega}{T_{e}}} - 1 \right)\Omega\, \delta_{-} \delta_{+} + \left( e^{-\left( \frac{\Omega_{-}}{T_{e}} + \frac{\Omega_{R}}{T_{d}} \right)} - 1 \right) \Omega_{-}\,\delta_{0}\delta_{-} + \left( e^{-\frac{\Omega_{+}}{T_{em}}} - e^{-\frac{\Omega_{R}}{T_{v}}} \right)\Omega_{+}\,\delta_{0}\delta_{+}   \right]
\label{current_dot2}	
\end{equation}
where
\begin{equation}
\mathcal{K} = \frac{1}{\Omega_R}\left[\left( 1 + e^{-\frac{\Omega_{-}}{T_{e}}} \right)\delta_{-} +  \left( 1 + e^{-\frac{\Omega_{+}}{T_{e}}} \right)\delta_{+} + \left( 1 + e^{-\frac{\Omega_{R}}{T_{d}}} \right) \delta_{0}\right]>0.
\label{current_dot3}	
\end{equation}
The expressions \eqref{current_dot1}-\eqref{current_dot3} can be used to analyze particular experimental settings. Here, we show only analytically the phenomenon of switching from the cooling to the heating regime observed in the experiment \cite{Vogl:2009}. We begin with the simplifying assumptions which are typically valid in quantum optics domain,
\begin{equation}
\Omega \gg T_{e}, \,  T_{d},\quad  \Omega_R \ll \Omega, \quad \Omega_R\ll T_{d}.
\label{conditions}	
\end{equation}
The last inequality holds for small detuning $\Delta = \omega_0 - \Omega$ only, but we are interested in the region where detuning changes its sign. Under the conditions \eqref{conditions} we can put
\begin{equation}
e^{-\frac{\Omega_{\pm}}{T_{e}}}\simeq 0, \quad e^{-\frac{\Omega_{R}}{T_{d}}}\simeq 1, \quad G^{e}(\Omega_{\pm})\simeq G^{e}(\Omega),
\label{conditions1}	
\end{equation}
and using \eqref{current_dot1}-\eqref{current_dot3} we obtain
\begin{equation}
\tilde{\mathcal{J}}^d \simeq  \mathcal{D}\, \Delta , \quad \mathcal{D}=\frac{\delta_{0}\, G^{e}(\Omega)}{2 \delta_{0} + \delta_{-} + \delta_{+} }> 0 ,
\label{current_dot4}	
\end{equation}
while
\begin{equation}
\tilde{\mathcal{J}}^e \simeq -\Omega\frac{2\delta_{-}\delta_{+}+ \delta_{0}(\delta_{-} + \delta_{+})}{2 \delta_{0} + \delta_{-} + \delta_{+}} < 0 ,
\label{current_em}	
\end{equation}
As a consequence, we obtain two regimes for small detuning $|\Delta|$:\\

(a) \emph{Cooling regime}: For red detuning (i.e., $\Omega < \omega_0$),  $\Delta > 0$ and $\tilde{\mathcal{J}}^d >0$, which means that heat flows from the dephasing bath to TLS. The system acts as a heat pump cooling the dephasing bath and heating the electromagnetic one at the expense of the work provided by the laser field.\\

(b) \emph{Heating regime}: For blue detuning (i.e., $\Omega > \omega_0$),  $\Delta < 0$, both heat currents $\tilde{\mathcal{J}}^d, \tilde{\mathcal{J}}^e < 0$, which means that both baths are heated at the expense of the laser field.
\section{Conclusions}
The presented formalism based on Markovian master equations describing periodically driven quantum open systems weakly coupled to heat baths is very flexible and possesses a large number of possible applications. It is consistent with thermodynamics and leads to  simple forms of kinetic equations, heat currents, and stationary power. This allows us to study, to a large extent analytically, various designs of microscopic machines, such as refrigerators and engines. The model briefly discussed here as Example 2 will be used in forthcoming papers for a detailed quantitative explanation of the experimental results of \cite{Vogl:2009} and as a basic ingredient of a solar engine model.

\section{Acknowledgements}

K.S. is supported by the Foundation for Polish Science TEAM project cofinanced by the EU European Regional Development Fund and D.G.-K.  by the CONACYT. R.A. acknowledges support by the Polish Ministry of Science and Higher Education via Grant No. NN 202208238.\\
\\
K.S. and D.G.-K. contributed equally to this work.


\begin{thebibliography}{10}

\bibitem{Sargent:2007}
P. Meystre and M. Sargent III,
\emph{Elements of Quantum Optics}, 4th ed.
(Springer, Berlin, 2007);
H. J. Carmichael, \emph{Statistical Methods in Quantum Optics 1} (Springer, Berlin, 2002).
\bibitem{Geva:1995}
E. Geva, R. Kosloff, and J. L. Skinner,
J. Chem. Phys. \textbf{102}, 8541 (1995).

\bibitem{Kamleitner:2011}
H-P. Breuer and F. Petruccione, Phys. Rev. A \textbf{55}, 3101 (1997); S. Kohler, T. Dittrich, and P. Hanggi, Phys. Rev. E \textbf{55}, 300 (1997); A. G. Kofman and G. Kurizki, Phys. Rev. Lett. 87, 270405 (2001); J. P. Pekola, V. Brosco, M. Mottonen, P. Solinas, and A. Shnirman,
Phys. Rev. Lett. \textbf{105}, 030401 (2010);
B. H. Wu, and C. Timm. Phys. Rev. B \textbf{81}, 075309 (2010); I. Kamleitner and A. Shnirman,
Phys. Rev. B \textbf{84},235140 (2011).

\bibitem{Davies:1974}
E. B. Davies, Commun. Math. Phys. \textbf{39}, 91 (1974).

\bibitem{Alicki:2006}
R. Alicki, D. A. Lidar and P. Zanardi, Phys. Rev. A {\bf 73}, 052311 (2006);
R. Alicki, D. Gelbwaser-Klimovsky, and G. Kurizki, arXiv:1205.4552v1 [quant-ph].

\bibitem{Levy:2012}
A. Levy, R. Alicki, and R. Kosloff,
Phys. Rev. E \textbf{85}, 061126 (2012); M. Kolar, D. Gelbwaser-Klimovsky, R. Alicki, and G. Kurizki, Phys. Rev. Lett. 109, 090601 (2012).

\bibitem{Alicki:1979}
R. Alicki, J. Phys. A \textbf{12}, L103 (1979); R. Kosloff, J. Chem. Phys. \textbf{80}, 1625 (1984), E. Geva and R. Kosloff, Phys. Rev. E \textbf{49}, 3903 (1994); R. Alicki, M. Horodecki, P. Horodecki, and R. Horodecki, Open Systems and Information Dynamics \textbf{11}, 205 (2004);
 D. Segal, and A. Nitzan, Phys. Rev. E \textbf{73}, 026109 (2006); E. Boukobza and D. J. Tannor, Phys. Rev. A \textbf{ 74}, 063823 (2006);
 N. Linden, S. Popescu, and P. Skrzypczyk, Phys. Rev. Lett. \textbf{105}, 130401 (2010); N. Erez, G. Gordon, M. Nest, and G. Kurizki, Nature, \textbf{452}, 724 (2008); J. Gemmer, M. Michel, and G. Mahler, \emph{Quantum Thermodynamics} (Springer, Berlin, 2010); B. Cleuren, B. Rutten, and C. Van den Broeck, Phys. Rev. Lett. \textbf{108},
120603 (2012); A. Mari and J. Eisert, Phys. Rev. Lett. \textbf{108}, 120602 (2012).

\bibitem{Spohn:1978}
H. Spohn, J. Math. Phys. \textbf{19}, 1227 (1978).

\bibitem{Vogl:2009}
U. Vogl, M. Weitz,  Nature \textbf{461}, 70 (2009);
U. Vogl, A. Sab, S. Habelmann and M. Weitz, Journal of Modern Optics \textbf{58}, 1300 (2011).

\bibitem{Xin}
Xin-Qi Li, H. Nakayama, and Y. Arakawa, Phys. Rev. B \textbf{59}, 5069 (1999).

\bibitem{Neubauer:2012}
A. Neubauer, S. Yochelis, I. Popov, A. Ben Hur, K. Gradkowski, U. Banin, and Y. Paltiel, J. Phys. Chem. C \textbf{116}, 15641 (2012).
\end{thebibliography}
\end{document}